# Hyperpolarization-enhanced NMR spectroscopy with femtomole sensitivity using quantum defects in diamond

**Authors:** Dominik B. Bucher,[1,2*] David R. Glenn,[1] Hongkun Park,[1,3,4] Mikhail D. Lukin,[1] Ronald L. Walsworth[1,2,4*]

**Affiliations:**
[1]Department of Physics, Harvard University, Cambridge, MA
[2]Harvard-Smithsonian Centre for Astrophysics, Cambridge, MA
[3]Department of Chemistry and Chemical Biology, Harvard University, Cambridge, MA
[4]Center for Brain Science, Harvard University, Cambridge, MA
*Correspondence to: rwalsworth@cfa.harvard.edu or dominik.bucher@cfa.harvard.edu

**Abstract:**
Nuclear magnetic resonance (NMR) spectroscopy is a widely used tool for chemical analysis and molecular structure identification. Because it typically relies on the weak magnetic fields produced by a small thermal nuclear spin polarization, NMR suffers from poor molecule-number sensitivity compared to other analytical techniques. Recently, a new class of NMR sensors based on optically-probed nitrogen-vacancy (NV) quantum defects in diamond have allowed molecular spectroscopy from sample volumes several orders of magnitude smaller than the most sensitive inductive detectors. To date, however, NV-NMR spectrometers have only been able to observe signals from pure, highly concentrated samples. To overcome this limitation, we introduce a technique that combines picoliter-scale NV-NMR with fully integrated Overhauser dynamic nuclear polarization (DNP) to perform high-resolution spectroscopy on a variety of small molecules in dilute solution, with femtomole sensitivity. Our technique advances mass-limited NMR spectroscopy for drug and natural product discovery, catalysis research, and single cell studies.

**Main text:**
Nuclear magnetic resonance (NMR) sensors based on nitrogen vacancy (NV) centers, point quantum defects in diamond, provide unprecedented detection of signals from small sample volumes[1–3]. While most early realizations of NV-detected NMR had limited spectral resolution (~1 kHz), recent work has shown that resolution ~1 Hz, sufficient to observe chemical shifts and scalar couplings ('J-couplings'), can be achieved in micrometer-scale NV-NMR detectors by employing a synchronized readout technique[4–6]. This advance opens the possibility of applying NV-NMR to a variety of next-generation analytic technologies, such as single-cell analysis[7] and metabolomics[8,9], and high-throughput screening of mass-limited chemical reactions[10–12]. However, because the relevant sample volumes are so small (picoliter-scale), NV-NMR spectroscopy has to date only been applicable to pure molecular samples[4,13]. This restriction precludes many potential chemical, biochemical, and biophysical applications, unless sensitivity improvements can be realized to enable the detection of dilute molecules in solution.

Here, we demonstrate a new technique to address this challenge using high-resolution, micrometer-scale NV-NMR in combination with *in-situ* hyperpolarization of the sample nuclear spins, resulting in an improvement of more than two orders of magnitude in molecule-number sensitivity for picoliter-scale sample volumes. The key innovation is to combine the NV-NMR with Overhauser dynamic nuclear polarization (DNP)[14–16] to transfer the thermal polarization of dissolved molecular radicals to the nuclei of sample molecules of interest. Integration of the Overhauser DNP system with the NV-NMR detector is technically straightforward, because the latter incorporates an efficient GHz-frequency antenna applicable both to NV spin manipulation and to DNP. The combined instrument provides a proton number sensitivity ~10 picomole/(Hz)$^{½}$, which enables high-resolution NMR spectroscopy from a variety of small molecules in solution at the scale of a single cell, with a sensitivity floor ~ 50 femtomole.

The NV-NMR spectrometer consists of a synthetic diamond chip, doped with a high concentration ($3 \times 10^{17}$ cm$^{-3}$) of NV centers in a thin (13 µm) layer at the diamond surface. The active area of the NV ensemble sensor is defined by a focused green laser beam (wavelength $\lambda = 532$ nm, spot diameter ~20 µm, see Figure 1a inset) used to initialize and read out the NV electronic spin states. This arrangement results in an effective NMR sensing volume of ~10 pL when a liquid sample is placed in contact with the diamond surface (Figure 1a)[4]. The laser is aligned for total internal reflection within the diamond to reduce the light intensity within the sample and minimize potential photobleaching of the dissolved molecular radicals. The diamond is oriented with the [111] axis parallel to the bias magnetic field ($B_0 = 84.7$ mT) provided by a feedback-stabilized electromagnet, and the resulting NV electron spin resonance transitions are driven by a wire loop antenna placed immediately above the diamond surface. Importantly, the bandwidth of this antenna is selected such that it may also be used to drive electron spin transitions in TEMPOL (4-Hydroxy-TEMPO) molecular radicals dissolved in the liquid sample. Experiments then proceed by applying alternating blocks of: (i) Overhauser DNP driving to the dissolved molecular radicals, to transfer thermal electron spin polarization to nuclear spins of the sample (Figure 1c); and (ii) detection of the sample's free nuclear precession (FNP) signal by a coherently averaged synchronized readout (CASR) magnetometry pulse sequence on the NV ensemble sensor [4] (Figure 1d). The FNP is induced by applying a π/2-pulse on the hyperpolarized nuclear spins of the sample.

**Results**

We first performed experiments to test the efficacy of DNP-enhanced NV-NMR using a sample of deionized water. TEMPOL radicals were dissolved in the water at a concentration of 20 mM, and experiments were carried out using a DNP Rabi frequency of 10 MHz (0 MHz) for the DNP (control) experiment. All other pulse sequence parameters were held constant. Comparison of the CASR-detected NMR spectra showed a ~230× increase in signal magnitude using DNP compared to the control experiment without DNP, consistent with the expected Overhauser DNP enhancement for TEMPOL at low magnetic field[16–18] (Figure 2a). To achieve this hyperpolarization enhancement of the FNP signal, we optimized polarization transfer from the electronic spins to the sample proton spins by recording the peak CASR-detected FNP signal amplitude while sweeping either the carrier frequency of the DNP drive (Figure 2b) or the DNP Rabi frequency (Figure 2c). In the first experiment, a triplet structure was visible in the CASR signal enhancement factor due to hyperfine splitting of the driven $^{14}$N electronic spin in the TEMPOL radical. In the second experiment, a maximum was observed in in the CASR signal enhancement at a Rabi frequency of ~10 MHz, saturating at higher power likely due to technical issues associated with sample heating and/or the microwave drive electronics. Addition of TEMPOL radicals to the sample resulted in a proton spin population lifetime of $T_1 \approx 150$ ms, well-matched to the operating linewidth of our NV-NMR sensor. Longer sample $T_1$ could be achieved by decreasing the TEMPOL concentration, with only a modest reduction in DNP signal enhancement (Figure S1). For a given initial TEMPOL concentration, the observed NMR signal enhancement remained constant over several days of experiments, indicating no appreciable decrease in concentration due to photobleaching.

We determined the molecule and proton number sensitivity achievable with DNP-enhanced NV-NMR spectroscopy in our system (Figure 3a) using samples of tert-butanol [$(CH_3)_3COD$, abbreviated t-BuOD] dissolved in heavy water ($D_2O$). The t-BuOD proton NMR spectrum is resolved by 3.55 ppm [19] (or ~13 Hz at $B_0 = 84.7$ mT) from residual semi-heavy water (HDO), which occurs in trace quantities in the solvent. By preparing samples with successive dilutions, we observed DNP-enhanced CASR signals from a sample size of ~ 50 femtomoles (molecule number, equivalent to a molecule concentration of 5.3 mM in the 10 pL detection volume) with a signal-to-noise ratio (SNR) of 3.5, after 5000 s of averaging (Figure 3b). This corresponds to a molecule number sensitivity of 3.2 pmol/Hz$^{½}$ for t-BuOD and a proton number sensitivity of 29 pmol/Hz$^{½}$, which is similar to the observed proton number sensitivity in hyperpolarized water of 10 pmol/Hz$^{½}$ (Figure 2a). In all cases the sensitivity is defined to a signal to noise (SNR) of 3. The quoted sensitivity includes time taken for both the hyperpolarization and FNP CASR signal detection components of the pulse sequence. To provide context for this result, we compare to reported sensitivities for several microscale inductive NMR detector technologies (Figure 3c). The inductive detectors operate at higher magnetic field (typically 4 – 14 T), but do not use hyperpolarization. Direct comparison indicates that DNP-enhanced NV-NMR provides superior number sensitivity and comparable concentration sensitivity to established inductive detection techniques, while also operating at smaller sample volume.

To investigate the generality of the approach, we performed DNP-enhanced NV-NMR spectroscopy on a variety of small molecules in solution (all at 0.8 M concentration). Samples of xylene [$(CH_3)_2C_6H_4$] dissolved in deuterated dimethyl sulfoxide (DMSO-$d_6$), dimethylformamide [$(CH_3)_2NC(O)H$] dissolved in $D_2O$, and trimethyl phosphate [$PO(OCH_3)_3$] dissolved in $D_2O$ were measured with CASR acquisition times of 50 seconds (Figures 4a-c). In each case, molecular NMR spectra were observed with SNR $\approx$ 25, and lineshape fits yielded the expected spectral parameters (line splittings and amplitudes) due to chemical shifts, J-coupling interactions, and relative proton abundances. The observed spectral linewidths were on the order of $\Delta f \approx 8 - 10$ Hz for each measurement. This was consistent with previously reported spatially-inhomogeneous linewidths in our NV-NMR spectrometer due to susceptibility-induced broadening[4], indicating that introduction of molecular radicals into the samples did not degrade system performance. Finally, we acquired DNP-enhanced NV-NMR spectra from a sample of the nucleobase thymine [$C_5H_6N_2O_2$] dissolved in DMSO-$d_6$ (Figure 4d). This measurement required an averaging time of 500 s to obtain an SNR of 20, largely because of a broadened resonance of the labile N-H protons, which we attributed to fast exchange with residual water in the solvent.

**Discussion**

Overhauser DNP using dissolved molecular radicals is an effective and technically straightforward hyperpolarization method to improve the sensitivity of NV-NMR spectroscopy at the micrometer-scale by more than two orders of magnitude. For proton NMR at the macroscopic scale, strong Overhauser signal enhancement has been demonstrated using inductive detectors for bias magnetic fields up to approximately ~1.5 T [17]. This suggests that at least one additional order of magnitude sensitivity enhancement is achievable in the present system. Nevertheless, increased bias fields up to 3 – 4 T are desirable to increase chemical shift resolution. While driving electron spin resonance transitions at such fields is technically challenging, successful demonstrations of both NV magnetic sensing[13,20] and Overhauser DNP[21,22] have been reported in the literature. Furthermore, the extremely small sample volumes accessible with the NV-NMR spectrometer may help to mitigate these challenges, due to both (i) reduced dielectric absorption by small samples, and (ii) the possibility of using small mode-volume resonators to efficiently drive the electronic spins.

We note that a variety of alternative hyperpolarization schemes have been proposed for NV-NMR sensors that take advantage of the optically-pumped NV electronic spins themselves as the hyperpolarization source[23–27]. However, while the achievable polarization of the NV centers is near unity, the small surface-to-volume ratio of the planar diamond chip detector geometry greatly limits the potential effectiveness of such methods in micrometer-scale sample volumes (details are available in the supplementary note 1). One proposed solution is to instead partially fill the sample volume with NV-doped nanodiamonds, overcoming geometric limitations of the planar diamond surface as a polarization source[23]. In this case, however, rotational freedom of the individual nanodiamond particles results in a random distribution of NV electronic spin orientations relative to the bias magnetic field, greatly complicating the procedure for polarization transfer[28]. Furthermore, it is often undesirable in practice to apply strong optical pumping to nanodiamonds within the sample volume, due to the possibility for photochemical effects to alter or degrade the sample. For these reasons, we find the introduction of molecular radicals into the solvent to be both an effective and practical hyperpolarization technique for NMR signal enhancement at the micrometer scale.

**Conclusion**

The ability to measure NMR signals with femtomole sensitivity from picoliter sample volumes will enable new ultra-sensitive and high-throughput analytics applications. For example, in drug development and natural products research, the current state of the art for large scale screens of binding affinity involves high-throughput nanomole-scale synthesis combined with mass spectrometry[10–12]. Introduction of ultrasensitive NMR spectroscopy to such a pipeline might simplify sample preparation and provide superior isomeric distinguishability. In the field of metabolomics, the excellent volume selectivity of NV-NMR may enable quantitative studies at the single-cell level[29,30]. Finally, while the present work has emphasized NV-NMR spectroscopy, we note that Overhauser DNP should be equally applicable to magnetic resonance imaging (MRI) techniques. In combination with strong pulsed field gradients[31], NV-detected MRI may enable studies of water diffusion and transport in cells and tissue at the micrometer scale[32].

## Methods

**NV ensemble NMR sensor.** The micrometer-scale NMR sensor is based on a $^{12}$C enriched (99.999%) chemical vapor deposition (CVD) diamond chip (2mm x 2mm x 0.5 mm) with a bulk nitrogen ($^{14}$N) concentration of <8.5 x $10^{14}$ cm$^{-3}$ (Element Six). The diamond is cut so that the lateral faces are perpendicular to [110] and the top face perpendicular to the [100] crystal axis. During the growth process, the CVD gas mixture was modified to generate a nitrogen-enriched surface layer of ~13 μm thickness, with $^{14}$N density of ~4.8 x $10^{18}$ cm$^{-3}$. Electron irradiation (flux of 1.3 x $10^{14}$ cm$^{-2}$ s$^{-1}$) for 5 h and subsequent annealing at 800 °C in vacuum yielded a dense NV ensemble (~3x$10^{17}$ cm$^{-3}$) in the nitrogen-enriched layer. The $T_2^*$ of the NV ensemble, measured using a Ramsey pulse sequence, is ≈750 ns, while the Hahn Echo time $T_2$ is ≈6.5 μs. All four diamond edges are polished at a 45° angle so that the top surface of the diamond is 1mm x 1mm. This geometry permits laser excitation of the NV centers using total internal reflection of the incident beam, which reduces light intensity at the sample position above the diamond. The 532 nm laser light is provided by a diode pumped solid state laser (Coherent Verdi G7), which is pulsed by an acousto-optic modulator (AOM) (IntraAction ASM802B47). Each pulse is 5 μs, where the first microsecond is used to readout the NV-state and the remaining time repolarizes the NV centers. The laser power is around 150 mW, focused down to a spot size of ~20 μm. The diamond is aligned so that the [111] axis is parallel to the external magnetic bias field ($B_0$). The ensemble NV magnetometer has a noise floor of ~20 pT/Hz$^{-1/2}$. Details about light collection, light detection, and the sample holder are described in Glenn et al. [4].

**Magnetic bias field.** The magnetic bias field, $B_0$ = 84.7 mT, is produced by an electromagnet (Newport Instruments Type A). At this field, the NV resonance frequency ($|m_s=0\rangle$ to $|m_s=-1\rangle$) is ≈500 MHz; the TEMPOL electronic spin resonance frequency is ≈2.37 GHz; and the proton NMR resonance frequency is ≈3.606 MHz. The bias field is stabilized with a second NV-diamond magnetometer (feedback sensor) as described in Glenn et al.[4]. For experiments longer than 5 minutes, slow drifts between the NV-NMR experiment and the feedback sensor are corrected every 5 minutes by measuring the magnetic field with the CASR sensor using an ESR frequency sweep. The microwave drive for the feedback sensor is delivered by a separate antenna, positioned immediately adjacent to the ESR sensor and driven independently from the main DNP-CASR experiment.

**DNP-CASR pulse sequence parameters.** The full DNP-CASR pulse sequence is divided in two parts: (a) the Overhauser hyperpolarization pulse sequence; and (b) FNP detection via a CASR readout sequence. Both are controlled by a programmable pulse generator (Spincore PulseBlaster ESR-PRO, 500 MHz). The Overhauser MW pulse duration is set to ~2 × NMR $T_1$ of the sample (typically around 2 × 150 ms). After the Overhauser sequence, a π/2 pulse with a duration of ≈150 μs is applied on the hyperpolarized proton sample to generate the FNP. The FNP signal is then read out with the CASR sequence, which is programmed on an arbitrary waveform generator (Tektronix AWG 7122C) and triggered by the pulse generator. The full CASR sequence duration is ~4 × NMR $T_2^*$ of the sample (typically around 4 × 50 ms). The CASR sequence (for details see [4]) is based on XY8-6 subsequences with a duration of 12.45 μs, chosen to be an integer multiple of: (i) the NV drive period ($1/f_{NV}$ = 1/500 MHz = 2 ns); (ii) the synchronized readout detection period (3320/12 ns); and (iii) the clock of the waveform generator ($1/f_{clock}$= 1/(12 GHz)). The $\pi$ and $\pi/2$ pulse durations used in the XY8-6 sequences are ≈60 ns and 30 ns, respectively. Every second pulse sequence is repeated with a 180° phase shift applied to the last $\pi/2$ pulse, in order to reject laser and MW noise by subtracting successive pairs of measurements. Thus, one data point of the CASR readout is recorded for two XY8-6 sequences (i.e., every 24.9 μs), with a total readout of 8000 points (199.2 ms). The duration of one full DNP-CASR experiment is ~500 ms, which includes the Overhauser sequence and FNP CASR detection.

**MW equipment.** Pulse sequences for driving the NV centers (resonance frequency 500 MHz) are directly synthesized, including both the carrier frequency and the pulse modulation, using an arbitrary waveform generator (Tektronix AWG 7122C), then amplified by a 100 W amplifier (Minicircuits ZHL-100W-52+). The Overhauser drive field (resonance frequency 2.37 GHz) is produced by a signal generator (SRS SG384), pulsed using a microwave switch (Minicircuits ZAWSA-2-50DR+) controlled by the programmable pulse generator, and amplified with a separate 100 W amplifier (ZHL-100W-242+). Both amplified MW drive fields (NV drive and Overhauser drive) are combined using a power combiner (ZACS-242-100W+) and sent to a loop antenna (see Bucher et al., Nature Protocols 2018, submitted). The loop has a diameter of 1 mm and is mounted immediately above the diamond. With this configuration, maximum Rabi frequencies are ~30 MHz for driving the NV centers and ~40 MHz for driving the TEMPOL electronic spins. To estimate the TEMPOL Rabi frequency, we performed a Rabi experiment, detected by the NV, on intrinsic electronic spins (dark spins) in the diamond lattice at g≈2 [33]. The MW power delivery and the DNP enhancement of the sample NMR signal varied somewhat in different experiments due to: (i) slightly different antenna orientations upon rebuilding the sensor mount; and (ii) different sample properties (e.g., heat conductance, microwave absorption). We typically used a relatively low NV Rabi frequency of 8.3 MHz for CASR experiments.

**NMR drive coils.** For applying the $\pi/2$ pulse on the protons we use a homemade resonant coil at ≈3.606 MHz with a quality factor (Q) of ~140. We typically achieve a proton Rabi frequency of ~4 kHz by driving this coil using our signal source (Rigol DG 1032) without amplification.

**Data analysis.** Each FNP signal measurement gives a time-series dataset consisting of 8000 data points from the CASR sequence. The first 20 data points (≈0.5 ms) are discarded because of artefacts (e.g., coil ringdown) associated with the $\pi/2$ pulse on the proton spins. For experiments in which signal-averaging is required, the averaging is performed in the time domain, before data are mean-subtracted. The time series data are zero-padded to a length of 20,000 points (498 ms), corresponding to the full duration of the combined pulse sequence (CASR + Overhauser). In addition, we filter the datasets by multiplying by an exponential filter function $e^{-t/\tau_f}$ with a time constant $\tau_f$. We used a time constant $\tau_f$ = 50 ms for the data shown in Figure 2a, and $\tau_f$ = 250 ms for the data in Figures 3a, b and Figure 4. After averaging and filtering, the time series datasets are then Fourier transformed in Matlab. In all plots, we show the absolute value of the Fourier transformation (CASR signal). Each experiment was performed at least 3 times to verify reproducibility. In most cases, the variability of the measured signal size is dominated by antenna alignment or sample concentration uncertainties, rather than the intrinsic sensitivity of the CASR readout.

The dataset of Figure 3a is fit to the sum of two modified Lorentzian functions:

$$F(f) = \frac{A_{t-BuOD} \cdot LW_{t-BuOD}}{\sqrt{(f - f_{t-BuOD})^2 + LW_{t-BuOD}^2}} + \frac{A_{HDO} \cdot LW_{HDO}}{\sqrt{(f - f_{HDO})^2 + LW_{HDO}^2}}$$

Here, A is the amplitude, LW the linewidth, and f the resonance frequency for t-BuOD and HDO, respectively. The amplitude $A_{t-BuOD}$ of the t-BuOD component of the spectrum is plotted in Figure 3b. In Figures 2b and c we plot the amplitude of the CASR signal against the swept experimental parameter. All amplitudes are normalized to a synthetic magnetic AC signal, generated by an external loop antenna positioned near to our diamond. In all cases we define the sensitivity as well as the proton number limit of detection (nLOD) for a signal-to-noise ratio (SNR) of 3.

**Samples.** As a hyperpolarizing agent, we used TEMPOL from Sigma Aldrich (catalog number 176141) without further modification. The samples p-xylene, trimethyl phosphate, N,N-dimethylformamide, thymine, and tert-butanol were purchased from Sigma Aldrich (catalogue numbers 296333, 241024, D4551, T0376 and 471712). The deuterated samples $D_2O$ and dimethyl sulfoxide-d6 (DMSO-d6) were obtained from Cambridge Isotope Laboratories, Inc (catalog number DLM-4-100 and DLM-10-10). In the data of Figure 3a we observe an HDO resonance line that is caused by (i) residual water in the purchased sample, and (ii) water vapor absorption during handling in the laboratory. Diluted samples were prepared by either weighing the sample or using microliter pipettes. Solvents and samples were not degassed in any of the experiments.

**Data Availability:** The data that support the findings of this study are available from the corresponding author upon reasonable request.
**Code Availability**: Custom software routines for analyzing the data presented in this study were written in Matlab. These Matlab scripts are available from the corresponding author upon reasonable request.

**Acknowledgments:** We would like to thank Matthew Rosen for initial help with the Overhauser polarization scheme. **Funding:** This material is based upon work supported by, or in part by, the U. S. Army Research Laboratory and the U. S. Army Research Office under contract/grant number W911NF1510548. D.B.B. was partially supported by the German Research Foundation (BU 3257/1-1).

**Author contributions**: D.B.B., D.R.G., and R.L.W. invented the Overhauser enhanced and CASR NV-diamond spectroscopy techniques. D.B.B., D.R.G., and R.L.W. designed the experiments and analyzed the data. D.B.B. modified the NV-NMR spectrometer for hyperpolarization and carried out the experiments. M.D.L, H.P., and R.L.W. conceived the application of NV diamond magnetometry to NMR detection at short length scales. All authors discussed the results and participated in writing the manuscript.

**Competing interests:** Authors declare no competing interests.

**Additional information:** Correspondence and requests for materials should be addressed to R.L.W. and D.B.B.

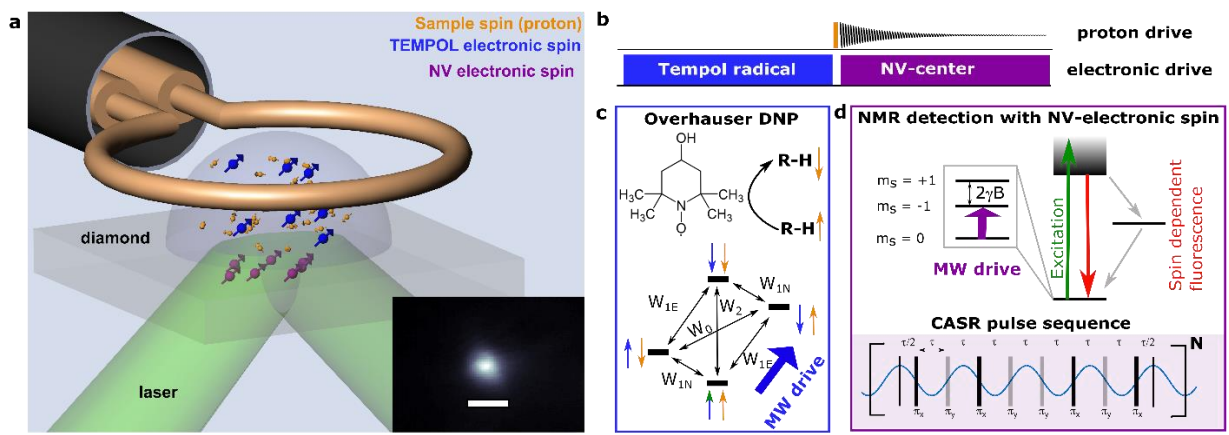

**Figure 1. NV-NMR spectroscopy with integrated hyperpolarization. a)** Experimental schematic. Microwave loop antenna near diamond chip drives both NV (purple) and TEMPOL electronic spins (blue). Hyperpolarized NMR signals from the sample nuclear spins (orange) are detected by NV ensemble fluorescence readout from the diamond chip. Inset: The sensor size is defined by the laser spot size on the diamond (scale bar is 30 μm). **b)** Integrated hyperpolarized NV-NMR spectroscopy pulse sequence. In the first half of the pulse sequence, the electronic drive is used to hyperpolarize proton spins in the sample via interactions with electronic spins in the TEMPOL radical using Overhauser dynamic nuclear polarization (DNP). In the second half of the pulse sequence, a π/2 pulse on the protons induces a free nuclear precession (FNP) signal from the hyperpolarized sample, which is detected by NV sensor spins via a coherently-averaged synchronized readout (CASR) pulse sequence. **c)** Overhauser DNP. Continuous microwave driving saturates the electronic spin transition of the TEMPOL radical. Relaxation leads to a net polarization of protons in an organic molecule (R-H), which is diffusing relative to the TEMPOL radical. **d)** FNP detection with NV-diamond and the CASR pulse sequence. The NV-center in diamond exhibits spin dependent fluorescence (right top) with triplet ground state spin transitions that can be accessed by resonant microwaves (left top). The proton FNP signal is detected by the CASR readout scheme (bottom), based on interspersed blocks of identical dynamic decoupling sequences synchronized to an external clock.

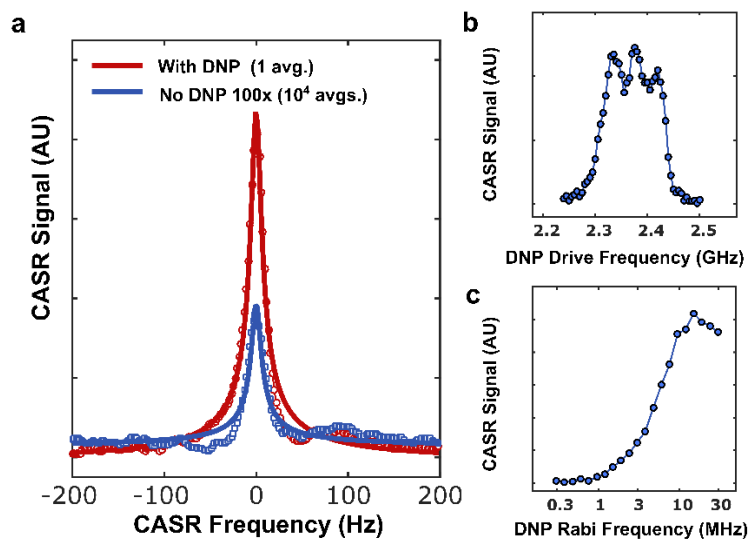

**Figure 2. Hyperpolarization-enhanced NV-NMR of water. a)** Comparison of NV-NMR spectra obtained from pure water with DNP (red circles, 1 spectrum averaged) and without DNP (blue circles, $10^4$ spectra averaged) using a coherently-averaged synchronized readout (CASR) pulse sequence. Line shape fits (solid lines) indicate a DNP signal enhancement of 230, with a proton number sensitivity of 10 pmol/Hz$^{½}$ for a signal to noise ratio (SNR) of 3. **b)** Magnitude of CASR-detected signal from a sample of pure water as a function of DNP drive frequency. The triplet structure arises from hyperfine coupling between the electron and $^{14}$N nuclear spins in the TEMPOL radical. **c)** Magnitude of CASR-detected signal from a sample of pure water as a function of DNP drive power, expressed in units of the electron Rabi frequency (see methods for details). The maximum DNP signal enhancement is reached at a drive Rabi frequency of approximately 10 MHz.

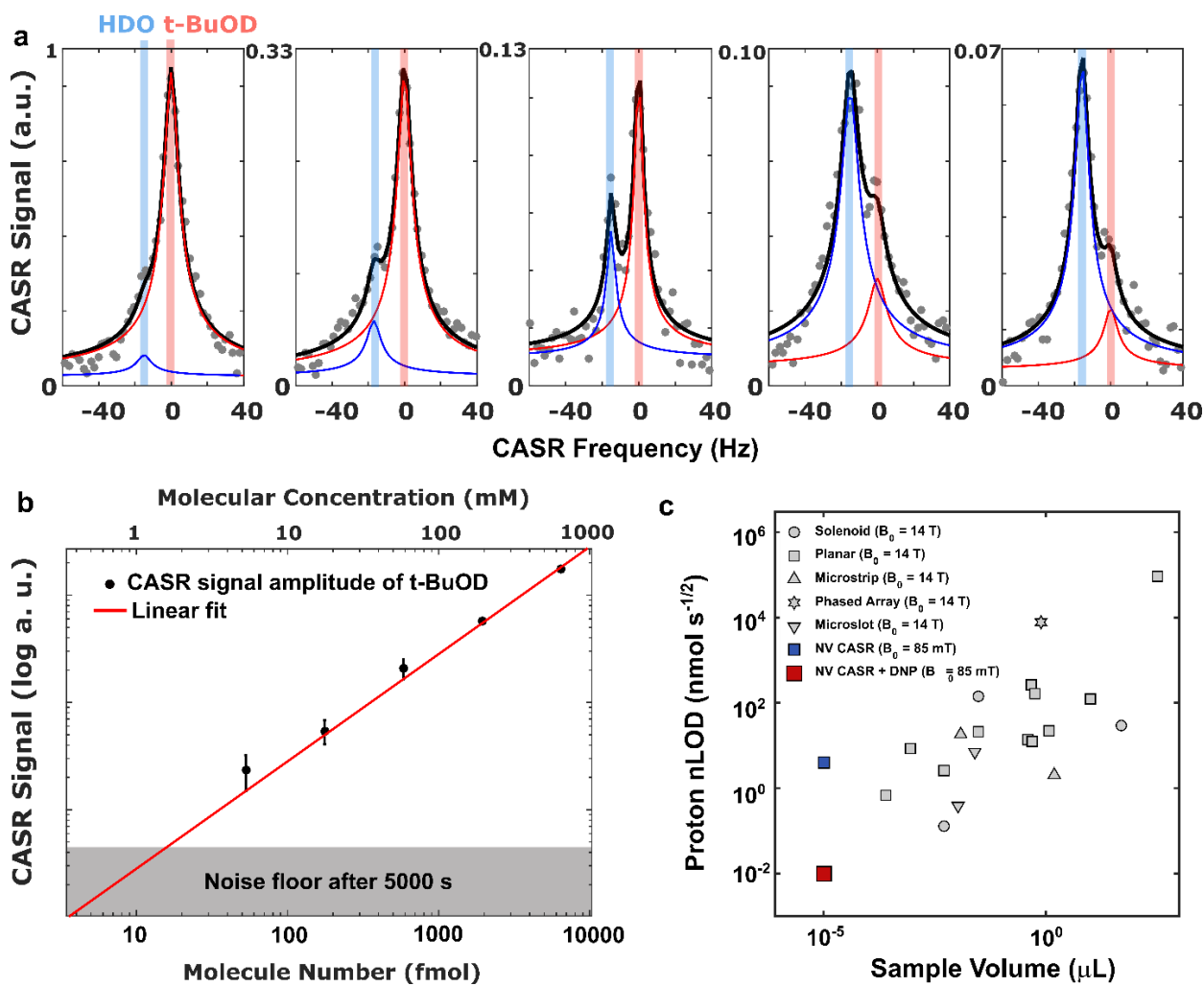

**Figure 3. Sensitivity of hyperpolarization-enhanced NV-NMR. a)** CASR-detected spectra of DNP-enhanced t-BuOD solutions at different millimolar concentrations (650 mM, 195 mM, 58.5 mM, 17.6 mM, 5.3 mM, from left to right) in $D_2O$. Spectra are fit to represent the t-BuOD sample (red solid line) and residual semi-heavy water (HDO) in the $D_2O$ (blue solid line), with red and blue vertical lines indicating the NMR resonance frequencies of t-BuOD and HDO, respectively. **b)** Plot of CASR signal fit amplitude of t-BuOD against number of sample molecules in the sensing volume (bottom axis) and sample molecular concentration (top axis). Grey area at the bottom marks the noise floor after 5000 s of averaging with a signal to noise ratio (SNR) ≈ 3.5 for a concentration of 5.3 mM, equivalent to ~50 fmol of sample molecules in the 10 pL detection volume. Error bars represent standard deviation (1σ) of CASR signal measured across three independent experiments, and are dominated by uncertainty in sample molecule number, which is larger than the measurement noise in each experiment. **c)** Proton number limit of detection (nLOD) comparison between NV-NMR, with and without DNP, and microscale inductive NMR detection technologies. Inductive sensitivities are scaled to a bias field of 14 T, whereas the NV-CASR + DNP sensitivity is reported at our typical operating bias field of ~85 mT. Microscale inductive sensor sensitivity data obtained from Badilita et al. [34]. The proton number limit of detection is defined for a SNR of 3.

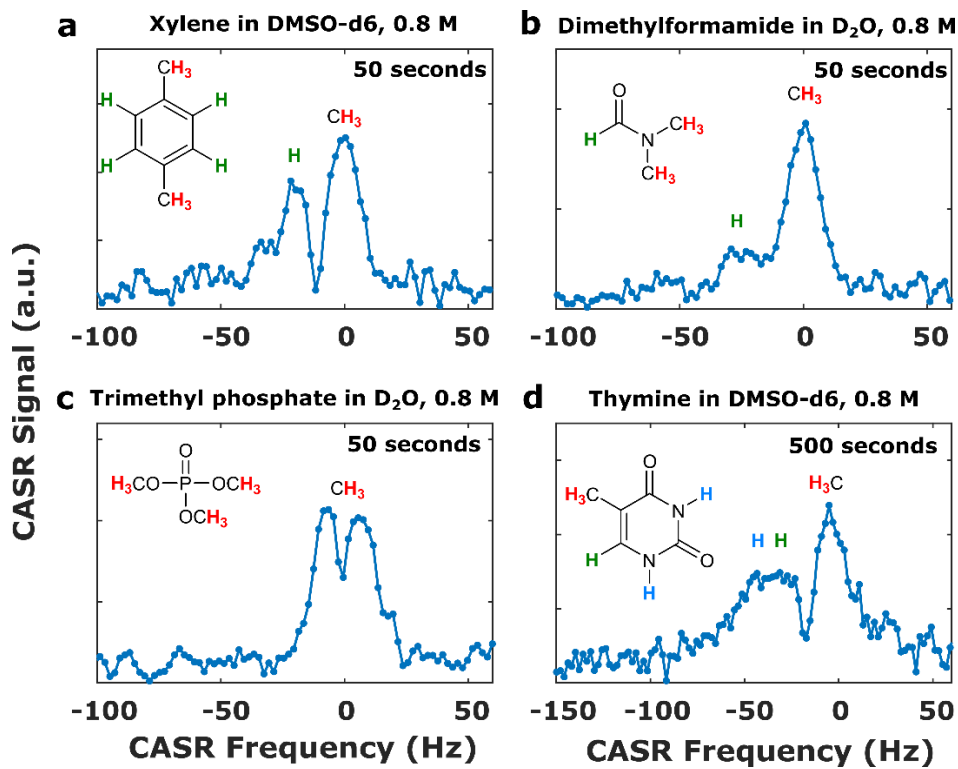

**Figure 4. CASR-detected NMR spectra of small organic molecules in solution. a)** Xylene in DMSO-d$_6$. **b)** Dimethyl formamide in D$_2$O. **c)** Trimethyl phosphate in D$_2$O. **d)** Thymine in DMSO-d$_6$. All samples were dissolved at a concentration of 0.8 M. The features of spectra in **a**, **b**, and **d** are dominated by chemical shifts, whereas in **c**, the J-coupling between $^{31}$P and $^1$H splits the CH$_3$ resonances into a doublet. In the thymine spectrum **d**, a line broadening effect is observable due to (N-H) proton exchange with small quantities of water dissolved in the DMSO-d$_6$. In all spectra, the frequency axis has been set to 0 Hz for the methyl resonance line.